\documentclass[11pt]{JHEP3}
\usepackage{amsmath}
\usepackage{amssymb}
\usepackage{xspace}
\usepackage[numbers]{natbib}
\usepackage[bottom]{footmisc}
\usepackage{relsize}
\usepackage{graphicx}
\usepackage{psfrag}
\usepackage{mathrsfs}

\DeclareMathOperator{\Tr}{Tr}

\DeclareMathOperator{\dn}{dn}
\DeclareMathOperator{\sn}{sn}
\DeclareMathOperator{\cn}{cn}

\newcommand{\adss}{$\text{AdS}_5\times \text{S}^5$\xspace}

\newcommand{\be}{\begin{equation}}
\newcommand{\ee}{\end{equation}}
\newcommand{\bea}{\begin{eqnarray}}
\newcommand{\eea}{\end{eqnarray}}
\newcommand{\su}{\alg{su}}
\newcommand{\sla}{\alg{sl}}

\newcommand{\sfrac}[2]{{\textstyle\frac{#1}{#2}}}
\newcommand{\half}{\sfrac{1}{2}}

\def \a {\alpha}
\def \pa{\partial}

\def\l{\lambda}
\def\eps{\epsilon}

\def\s{\sigma}

\def \pa{\partial}
\newcommand{\ellK}{{\rm K}}

\newcommand{\ellE}{{\rm E}}

\def\L{\mathscr{L}}
\def\bP{\bar{\Psi}}

\def\bl{\bar{\hskip 0.3mm\lambda}}
\def\l{\lambda}

\newcommand{\alg}[1]{\mathfrak{#1}}

\def \st {{\rm t}}
\def \lle {{\ell}}

\def \cE {{\cal E}}

\title{Linking B\"acklund  and Monodromy Charges \\ for Strings on \adss}

\author{Gleb Arutyunov$^1$ and Marija Zamaklar$^2$ \\
\llap{$^1$\,}Institute for Theoretical Physics\\
Utrecht University \\
Leuvenlaan 4, 3508 TD Utrecht\\ The Netherlands\\
~\\
\llap{$^2$\,}Max-Planck-Institut f\"ur Gravitationsphysik\\
Albert-Einstein-Institut\\
Am M\"uhlenberg 1\\ D-14476 Golm, Germany\\

\email{g.arutyunov@phys.uu.nl, marija.zamaklar@aei.mpg.de}}
\preprint{AEI-2005-003\\
SPIN-05/11\\ ITP-05/13\\hep-th/0504144}

\abstract{ We find an explicit relation between the two known ways of
generating an infinite set of local conserved charges for the string
sigma model on \adss: the B\"acklund and monodromy approaches. We
start by constructing the two-parameter family of B\"acklund
transformations for the string with an arbitrary world-sheet
metric. We then show that only for a special value of one of the
parameters the solutions generated by this transformation are
compatible with the Virasoro constraints.  By solving the B\"acklund
equations in a non-perturbative fashion, we finally show that the
generating functional of the B\"acklund conservation laws is equal to
a certain sum of the quasi-momenta. The positions of the quasi-momenta
in the complex spectral plane are uniquely determined by the real
parameter of the B\"acklund transform. }

\begin{document}

\section{Introduction and Summary}

New important insights into the conjectured duality between gauge
and string theories~\cite{M} have been gained in the last two
years. Although this duality relates a weakly coupled gauge theory
to a strongly coupled string theory in the \adss background, it
has been realized that certain (plane-wave) string states do admit
a direct comparison with composite operators of the ${\cal N}=4$
SYM \cite{BMN}. Furthermore, extending the semi-classical approach
of \cite{GKP}, a large sector of rotating multi-spin string
solutions has been found \cite{FTa}. These string solitons are
naturally described by a simple finite-dimensional integrable
system \cite{AFRT,ART} and they probe the structure of the
space-time beyond the plane-wave limit. From mathematical point of
view these solitons are the simplest examples of more general
finite-gap solutions of the classical string sigma model
\cite{KMMZ}. The related progress on the gauge theory side has
been based upon understanding the integrable properties of the
dilatation operator at leading \cite{MZ} and higher orders of
perturbation theory \cite{BKS}. This nicely generalizes and
extends the integrable structures found in QCD \cite{BDM}.

\medskip

The observed integrability of the \emph{classical} string sigma
model and the integrability of various spin chain Hamiltonians
emerging from the dilatation operator in the gauge theory open up
a new avenue to address the issue of string quantization in the
\adss background.\footnote{While in the non-planar sector
integrability is generically broken, it seems that for certain
processes, integrable structures might remain
preserved~\cite{Peeters}.} The necessity of understanding the
string spectrum is especially sharpened in the light of recently
observed discrepancies between the gauge and string theory
calculations~\cite{SS,Callan,Min}.  Although these mismatches can
presumably be attributed to the order in which limits are taken in
the string and gauge theories, the only way to fully resolve this
issue is to find the exact string spectrum in \adss and compare it
to that of gauge theory.

\medskip

At present it is unknown how to promote the observed
\emph{classical} integrability of the sigma model to the quantum
level. Inspired by the findings in the gauge theory~\cite{SS,BDS},
one can make an educated guess~\cite{AFS} for the \emph{quantum}
version of the classical Bethe equations of the string sigma
model~\cite{KMMZ}. However, the quantum string Bethe equations
describe the so-called $\su(2)$ subsector of the theory and they
are asymptotic, \emph{i.e.}~they require the \mbox{R-charges} of
the string states to be large. It is not yet clear to which extent
these equations can reproduce the full string spectrum.  Recently
these equations were extended in a beautiful way to the other
subsectors, $\sla(2)$ and $\su(1,1)$, and further intriguing
relations to the gauge theory quantities were found~\cite{S}.

\medskip

Success of string quantization crucially depends on the choice of
dynamical variables. Since it is not clear what kind of gauge is
most suitable for the quantization, it is important to extract
information in a covariant manner as much as possible.  Even
classically, restricting oneself to a particular gauge may
simplify some computations, but can make other computations
extremely difficult.  For instance, the uniform gauge of~\cite{AF}
is convenient for the construction of the perturbative expansion
(in the inverse powers of curvature) of the string Hamiltonian
around the plane-wave limit. However, this gauge is not suitable
to reach the flat space limit; here the AdS light-cone
gauge~\cite{MT} is appropriate. Thus, the best option would be to
explore as far as possible the classical/quantum integrability of
the string sigma model in a covariant manner.
\medskip

The integrability of the classical sigma model is manifested through
an infinite set of (commuting) conserved charges\footnote{There is
also the issue of the non-abelian symmetry which has been investigated
in~\cite{BPR}.}. Two ways to construct these charges\footnote{The
principal sigma model and its reductions also admit local \emph{higher
spin} conserved currents~\cite{Polyakov}. It would be interesting to
clarify their meaning in the context of the AdS/CFT duality
conjecture.} are known. One of them is through the so-called
B\"acklund
transformations~\cite{Pohlmeyer:1975nb,Ogielski:1979hv}. The other is
based on the fundamental linear problem and the associated monodromy
matrix~\cite{Zakharov:1973pp} (see also~\cite{FT} for a comprehensive
review). Recently, both of these methods have been used successfully
to reveal a close relation between the integrable structures of gauge
and string theories~\cite{AS,KMMZ}. More precisely, in~\cite{AS} the
generating function for the B\"acklund conservation laws associated
with the classical bosonic string sigma-model in the \emph{conformal
gauge} has been constructed. It has been evaluated on the so-called
rigid string solutions and furthermore shown to match (at two leading
orders of perturbation theory) with the generating function of the
commuting charges obtained on the gauge theory side (see
also~\cite{Enq}). Further interesting developments in this direction
include linking the B\"acklund transform with the geometric~U(1)
symmetry of the string phase space~\cite{Mikhailov:2004au}.

\medskip
On the other hand, in~\cite{KMMZ} the monodromy approach has been
used to classify the finite-gap solutions of the classical bosonic
string sigma-model. Here the spectrum of the model appears to be
encoded in integral equations of the Bethe type which also exhibit
an agreement with the Bethe equations describing the spectrum of
long operators in the Yang-Mills theory~\cite{KMMZ}. Recently,
this approach has been extended to the supersymmetric string
sigma-model~\cite{Beisert:2005bm} (see also~\cite{Alday:2005gi} on
related issues) and furthermore used to show an agreement with the
spectrum of the one-loop gauge theory.

\medskip
The features discussed above all point out that the two approaches of
studying strings on \adss --- the B\"acklund and the monodromy
approaches --- are undoubtedly related.  Surprisingly, we have not
found a simple explanation of this fact in the existing
literature. The main aim of the present paper is therefore to
understand the precise relation between these two, apparently
different approaches. Let us now describe the content of the paper and
the results obtained.

\medskip

In view of the importance of the covariant approach, we start by
constructing the covariant form of the B\"acklund transformations.
The B\"acklund transformations transform a solution of the
second-order evolution equations $X_0$ (the \emph{reference}
solution) into a new solution~$X(\lambda,x)$ (the \emph{dressed}
solution).  We identify a family of such transformations which
depend on two continuous (spectral) parameters, $\lambda$ and~$x$.
These transformations are defined in an arbitrary but \emph{fixed}
world-sheet metric~$\gamma$. In string theory, in addition to the
dynamical equations for the string embedding coordinates, one also
has the Virasoro constraints. These are just the equations of
motion for the world-sheet metric and once the field
configuration~$X$ is known, the metric is fully fixed. Since in
our construction we \emph{assume} that the reference and dressed
solutions have \emph{the same} world-sheet metric, one also has to
check that the new solution is compatible with this fixed metric
(\emph{i.e.}~one has to check that the new solution satisfies the
Virasoro constraints).\footnote{It would also be interesting to
understand whether the whole construction of the B\"acklund
transform can be carried out without assuming the invariance of
the world-sheet metric but requiring the reference and dressed
solutions to obey the Virasoro constraints.}  We show that this is
not always the case, and that this requirement introduces a
restriction on the value of the spectral parameter. By computing
the two-dimensional stress tensor for the dressed solution,
assuming the invariance of the world-sheet metric under the
B\"acklund transform, we find that it vanishes only for special
values of the spectral parameter~$\lambda$, namely for~$\lambda =
\pm1$. Thus, not every solution of the B\"acklund transform is
compatible with the conformal constraints. It seems that this
issue plays only a minor role in the general construction of
integrable models, but it is crucial for applications in string
theory.

\medskip

Given a periodic solution to the B\"acklund equations one can
construct an infinite set of conserved currents characterizing the
reference solution.  Building on the approach of Hanrad et
al.~\cite{HSS}, we solve the B\"acklund equations in a
non-perturbative fashion. In this process the B\"acklund equations are
reduced to the fundamental linear problem~\cite{Zakharov:1973pp} and,
therefore, their solutions are expressed in terms of the wave function
which solves the linear problem.  Using this explicit solution, we
rewrite the generating function of the B\"acklund conservation laws
via the eigenvalues of the monodromy matrix associated to the linear
problem. It turns out that the B\"acklund generating function is equal
to a certain sum of the logarithms of eigenvalues (quasi-momenta) of
the monodromy matrix; the positions of the individual quasi-momenta in
the complex plane are uniquely determined by the real spectral
parameter~$x$ of the B\"acklund transform. This establishes a direct
relation between the B\"acklund and monodromy approaches for strings
in \adss.

\medskip
In summary, we have discovered a very simple relation between two
apparently different approaches to the construction of an infinite set
of conserved charges for integrable models.  In the context of string
theory, our result could be used to shed some light on the formidable
problem of quantizing strings on \adss.  In particular, it seems to
suggest that the ``quantization'' of the classical Bethe equation
\cite{AFS}, originating from the monodromy approach, could equally
well be described in terms of the yet unknown \emph{quantum B\"acklund
transform}; the latter should be understood as quantization of the
classical B\"acklund equations.  To see whether this is really the
case for the full string theory on \adss, it would be important to
first understand whether and how the relations which we have
discovered for the bosonic string prevail once the fermions are taken
into account, \emph{i.e.}~for the classical Green-Schwarz superstring
on \adss. This question will be discussed elsewhere.

\medskip
For the convenience of the reader, let us summarize the
organization of the paper. In section~2 we introduce and describe
the general properties of the B\"acklund transform. In section~3
we determine the general requirements for the B\"acklund transform
to be compatible with the Virasoro constraints. In the next
section we discuss a general solution of the B\"acklund equations
in terms of the fundamental linear problem. Furthermore, in
section~5 we find a relation between the B\"acklund and the
monodromy conservation laws. In section~6 we present an
independent perturbative check of our basic formula relating the
B\"acklund and the monodromy charges. Finally, in two appendices
attached, we discuss the perturbative construction of B\"acklund
charges and also give some details on the gamma-matrix algebra.

\section{Preliminaries}

In this section we set up the notation and review some background
material which will be necessary for the derivation of the covariant
form of the B\"acklund equations.  The starting point is the (bosonic)
part of the sigma model action for strings in the \adss background,
\begin{equation}
\label{e:action}
\begin{aligned}
&{\rm S} = {
1\over 2}\int {\rm d}\tau {\rm d} \sigma~ \gamma^{\alpha \beta}
\Tr \left( \partial_\alpha g g^{-1}\partial_\beta g g^{-1} \right)
\, ,
\end{aligned}
\end{equation}
where the indices $\alpha,\beta=(\tau, \sigma)$ refer to the
world-sheet time and space directions. Here the matrix~$g$ describes
an embedding of the \adss space into the group ${\rm
SU}(2,2)\times {\rm SU}(4)$  and $\gamma^{\alpha \beta}=\sqrt{-h}
h^{\alpha \beta}$ is the Weyl invariant tensor constructed from the
Lorentzian world-sheet metric $h_{\alpha \beta}$, and it has
$\det\gamma=-1$. The string tension in~(\ref{e:action}) is set to
unity. The equations of motion for the dynamical fields~$g,
\gamma^{\alpha \beta}$ derived from the action~(\ref{e:action})
are
\begin{align}
\label{e:eom-q}
&\partial_\alpha ( \gamma^{\alpha \beta} \partial_\beta g g^{-1})
=
\partial_\alpha ( \gamma^{\alpha \beta} g^{-1}  \partial_\beta
g)=0\, ,
\\
\label{e:eom-h} &\Tr(\partial_\alpha g g^{-1}\partial_\beta g
g^{-1})- {1\over 2} \gamma_{\alpha \beta}\,  \Tr(\partial_\rho g
g^{-1}
\partial_\delta g g^{-1}) \gamma^{\rho \delta}   = 0 \, .
\end{align}
Equations (\ref{e:eom-q}) are the conservation laws
for the left, $A_{\rm L}$, and the right, $A_{\rm R}$, currents
\begin{equation}
\begin{aligned}
A_{\rm L}^{\alpha}=  \gamma^{\alpha \beta} \partial_\beta g
g^{-1}\, , ~~~~~~A_{\rm R}^{\alpha}= \gamma^{\alpha \beta} g^{-1}
\partial_\beta g \, .
\end{aligned}
\end{equation}
In what follows we will mainly use $A_{\rm L}$ and therefore to
save notation we will drop the subscript ${\rm L}$.

\medskip
We will assume that the group element $g$ has the block-diagonal
structure
{\small
\begin{eqnarray}
\label{e:parametrisation} g &=& \left(\begin{array}{cc}
   g_{\rm a} & 0 \\
   0  & g_{\rm s}
\end{array}\right)  \, ,
\end{eqnarray}
} where the matrices $g_{\rm a}$ and $g_{\rm s}$ belong to  ${\rm
SU(2,2)}$ and ${\rm SU(4)}$ respectively. To describe an embedding
of the \adss space into ${\rm SU(2,2)}\times {\rm SU(4)}$ it is
convenient to choose the following parametrization for the group
elements \cite{ART} \bea \label{embed} g_{\rm a}=p_{\rm
}^i\Gamma^i_{\rm a} \, , ~~~~~~~~g_{\rm s}=q_{\rm }^i\Gamma^i_{\rm
s}\, .\eea Here $i=1,\ldots, 6$. The $4\times 4$ gamma-matrices
$\Gamma^i_{\rm a}$ and $\Gamma^i_{\rm s}$ realize the {\it chiral}
representations of ${\rm SO(4,2)}$ and ${\rm SO(6)}$ respectively.
We summarize some of  their properties in appendix B. The
variables $p^i$ and $q^i$ parametrize the AdS space and the
five-sphere and they obey the constraints $q^iq^i=1$ and
$\eta_{ij}p^ip^j=-1$, where $\eta_{ij}$ has AdS signature.

\medskip

Before solving the Virasoro constraints (\ref{e:eom-h}), the
block-diagonal structure of~$g$ implies that the B\"acklund
transformations and the conservation laws associated to the AdS
and sphere sectors of the model are completely independent. Thus,
it is sufficient to discuss the corresponding theory for the
sphere part of the model; extension to the AdS sector goes without
any difficulty. Therefore in what follows we set 
\bea g\equiv
g_{\rm s} 
\eea 
confining our explicit treatment of the B\"acklund theory to the
sphere case.

\medskip

Finally we have to take into account the Virasoro constraints
(\ref{e:eom-h}) which express the condition of vanishing of the
two-dimensional stress tensor. Given a solution $g$ of
equations~(\ref{e:eom-q}), equation (\ref{e:eom-h}) can be solved for the
world-sheet metric $\gamma^{\alpha \beta}$:
\begin{equation}
\label{e:gamma} \gamma_{\alpha \beta} = {\rm \theta_{\alpha \beta}
\over \sqrt{-{\rm det} \, \theta_{\alpha \beta}} } \, , \quad
\quad {\rm \theta}_{\alpha \beta} = \Tr(A_{\alpha}A_{\beta}
) \, .
\end{equation}
Thus, the AdS and sphere sectors of the model are related through
the Virasoro constraints. Generically the world-sheet metric
appears to be a function of both the AdS and sphere coordinates.
On the other hand, our definition of the B\"acklund transform (see
the next section) implies that the reference and dressed solutions
have the same wold-sheet metric~(\ref{e:gamma}). A priori such a
definition is not necessarily compatible with the Virasoro
constraints and below we will find further restrictions on the
B\"acklund transform which guarantee that this is indeed the case.

\section{The  B\"acklund Transformations and Conservation Laws}
Given any two solutions of the equations of motion, $g$ and
$\tilde{g}$, we can construct two conserved currents
$\tilde{A}_{\a}$ and $A_{\a}$. Let us now require the difference
of these quantities to be a topological current which is therefore
trivially conserved:\footnote{The currents $\tilde{A}_{\a}$ and
$A_{\a}$ are periodic functions of $\sigma$. This implies for
$\chi$ that $\chi(\sigma+2\pi)=\chi(\sigma)+{\rm const}$. However,
if we require coincidence of the corresponding Noether charges,~$\chi$
must be periodic.} 
\bea \label{start}
\gamma^{\a\beta}\tilde{A}_{\beta}-\gamma^{\a\beta}A_{\beta}=\epsilon^{\alpha\beta}\pa_{\beta}\chi\,. 
\eea 
Here we assume that the indices of both $A$ and $\tilde{A}$
are raised and lowered with a one and the same world-sheet metric
$\gamma$. The matrix $\chi$ depends on $\tilde{g}$ and $g$, and it
becomes constant when $\tilde{g}=g$. Subtracting from
equation (\ref{start}) its hermitian conjugate and using the fact that
$A$ and $\tilde{A}$ are anti-hermitian we obtain 
\bea \label{pchi}
\chi+\chi^{\dagger}={\rm C}\, , 
\eea 
where ${\rm C}$ is a constant matrix. The matrix $\chi$ must be also
invariant under the global transformation $g\to gh$ and should
transform as $\chi\to h\chi h^{-1}$ under the global rotations $g\to
hg$. In particular, equation (\ref{pchi}) will remain invariant under
these symmetry transformations provided we choose ${\rm C}$ to be
proportional to the identity matrix. Obviously, any $\chi$ with such
properties can be constructed in terms of a unitary matrix \bea
U=\tilde{g}g^{-1} \eea or its inverse. To restrict possible choices
for $\chi$ we therefore have to impose certain conditions on $U$
which would allow one to express all higher powers of $U$ or $U^{-1}$
in terms of~$U$. The simplest possibility is to take $\chi=\lambda U$,
where $\lambda$ is a complex (spectral) parameter. Since a phase of
$\lambda$ can always be absorbed by redefining $U$ we may assume that
$\lambda$ is real. In appendix B we will verify that reality of
$\lambda$ is compatible with our definition of the coset model. With
this choice we have \bea \label{char} U+U^\dagger=2\frac{x}{\lambda}
\mathbb{I}\, , \eea where $x$ and $\lambda$ are real numbers.  Hence,
in what follows we assume that the difference of the Noether currents
is of the form \bea \label{mback} \tilde{A}_{\a}- A_{\a}=\lambda\,
\epsilon_\alpha{}^\beta\, U_{\beta} \, \eea where we defined $
\epsilon_\alpha{}^\beta \equiv \, \gamma_{\alpha \delta} \,
\epsilon^{\delta \beta}$ and also $U_{\beta}\equiv \pa_{\beta}
U$. Obviously equation (\ref{mback}) represents a non-trivial
condition on ${\tilde g}$. We will refer to $\tilde{g}$ as the
B\"acklund transform of $g$.

\medskip

Since $U$ is unitary it can be diagonalized with a proper
unitary matrix. Then equation~(\ref{char}) allows one to determine the
eigenvalues of $U$. The eigenvalues appear to be degenerate -- two
of them are equal to $\ell$ and the other two to its complex
conjugate, $\bar{\ell}$, where \bea \label{re}
\ell=\frac{x}{\lambda}-i\sqrt{1-\left(\frac{x}{\lambda}\right)^2}
\, . \eea Moreover, the eigenvalue problem imposes a restriction
on the spectral parameter $x$: \bea -\lambda \leq  x \leq
\lambda\, . \eea

\medskip

\noindent The current $\tilde{A}_{\a}$ can be written via $U$ as
follows \bea \label{gt}
\tilde{A}_{\a}=\pa_{\a}UU^{-1}+UA_{\a}U^{-1}\, . \eea Thus, we see
that the B\"acklund transformation is in fact a certain {\it gauge
transformation}.\footnote{See \cite{Babelon:1993yd} on the
relation of B\"acklund transformations with the theory of
Poisson-Lie groups and dressing symmetries.} The basic relation
(\ref{mback}) can be written as the differential equation for the
matrix $U$: \bea \label{mback1}
U_{\beta}(\delta_\alpha{}^{\beta}-\lambda \epsilon_\alpha{}^\beta
U)=[A_{\a},U]\, . \eea
Using equation (\ref{char}) the last equation can be brought to the form
\bea \nonumber \kappa U_{\a}&=&-2x\lambda
A_{\a}+(1+\lambda^2)A_{\a}U-(1+\lambda^2-4x^2)UA_{\a}
-2x\lambda UA_{\a}U +\\
\label{Riccati} &+& \epsilon_\alpha{}^\beta
\Big(-\lambda(1+\lambda^2)A_{\beta}+2x A_{\beta}U+2x\lambda^2
UA_{\beta}-\lambda(1+\lambda^2)UA_{\beta}U \Big)\, , \eea where
$\kappa=(1+\lambda^2)^2-4x^2$. This is a matrix differential
equation of the \emph{Riccati type}; its solutions depend on two
spectral parameters $x$ and $\lambda$.

\medskip
Equation (\ref{Riccati}) implies an infinite number of
conservation laws. Indeed, define the following current \bea
\label{CC}
J^{\a}=\frac{1+\lambda^2}{\kappa}\Big[(1+\lambda^2)\gamma^{\a\beta}{\rm
Tr}(A_{\beta}U)+2x \eps^{\a\beta} {\rm Tr}(A_{\beta}U) \Big]\,
.\eea Using the Riccati equation (\ref{Riccati}), the equations of
motion for $A_{\a}$ and the zero-curvature condition
$\pa_{[\a}A_{\beta]}=[A_{\a},A_{\beta}]$ one can easily prove that
$\partial_{\a}J^{\a}=0$. The normalisation of the current is chosen
for later convenience. Assuming the solution $U$ to be a periodic
function of $\s$, $U(\s+2\pi)=U(\s)$, the generating function of
conserved charges is obtained by integrating the $\tau$-component
of the current:
\begin{equation}
\label{QJ}
\begin{aligned}
&{\rm {\bf Q}}(x,\lambda) = \int_0^{2\pi} {{\rm d} \sigma\over 2
\pi} J^{\tau}(\sigma) \, .
\end{aligned}
\end{equation}
Upon expansion over the spectral parameters the function ${\rm
{\bf Q}}(x,\lambda)$ generates an infinite set of integrals of
motion. It is worth stressing that both the B\"acklund
transformations and the conservation laws are determined for an
arbitrary world-sheet metric.

\medskip
We further notice that if we define  $\mathscr{L}$ as \bea
\label{fL}
\mathscr{L}_{\a}=\frac{1+\lambda^2}{\kappa}\Big((1+\lambda^2)A_{\a}+2x\eps_{\a}{}^{\beta}A_{\beta}\Big)
\eea then $\mathscr{L}$ satisfies the zero curvature condition.
The conserved current (\ref{CC}) takes a very simple form \bea
\label{simple} J^{\a}=\gamma^{\a\beta}{\rm
Tr}(\mathscr{L}_{\beta}U)\, . \eea More generally, introducing a
connection $\L(a,b)$ parametrized by the coefficients $a$ and $b$
\bea \label{Lc} {\L}_{\a}(a,b)=a
A_{\a}+b\eps_{\a}{}^{\beta}A_{\beta} \eea one can check that it
has zero curvature provided \bea a^2-b^2-a=0\, . \eea We will
refer to such a connection as the $\L$-operator.

Finally we note that the Riccati equation (\ref{Riccati}) can be
expressed in terms of the $\mathscr{L}$-operator (\ref{fL}) only
\begin{equation}
\label{Ri} (1+\lambda^2)U_\alpha = [{\mathscr L}_ \alpha, U] -
\epsilon_\alpha{}^\beta \left(\l {\mathscr{L}}_\beta + \l U
{\mathscr{L}}_\beta U-2xU\L_{\beta} \right) \, .
\end{equation}

\section{The Stress Tensor}

The definition of a conserved current requires a world-sheet metric.
In our construction of the B\"acklund transform we assumed that the
conserved currents $A^{\alpha}$ and $\tilde{A}^{\a}$ are defined with
one and the same world-sheet metric $\gamma^{\a\beta}$. As we have
already discussed in the introduction, in string theory the
Weyl-invariant metric $\gamma^{\a\beta}$ is fully determined by the
Virasoro constraints $T_{\a\beta}=0$, where $T_{\a\beta}$ is the
two-dimensional stress tensor. Suppose we are given a pair $g$ and
$\gamma^{\a\beta}$ which solves both the dynamical equations for $g$
and the Virasoro constraints. An important question we want to address
here is what are the general conditions on the B\"acklund solution
$\tilde{g}$ so that $\tilde{g}$ still solves the Virasoro constraints
with the same metric $\gamma^{\a\beta}$.  In other words, we require
vanishing of the stress-energy tensor~$T_{\a\beta}(\tilde{g},\gamma)$
for the B\"acklund solution.

\medskip
To elaborate on this issue, we first compute $\delta
\theta_{\a\beta}={\rm
Tr}(\tilde{A}_{\a}\tilde{A_{\beta}}-A_{\alpha}A_{\beta})$. Using
equation~(\ref{char}) it is easy to see that $U_{\a}$ obeys the
following equation \bea U_{\a}=UU_{\a}U \, .\eea This equation
together with equation (\ref{gt}) leads to
 \bea
\delta\theta_{\a\beta}={\rm
Tr}\Big(U_{\a}UA_{\beta}+U_{\beta}UA_{\a}+U_{\a}U_{\beta} \Big)\,
. \eea Furthermore, we use the B\"acklund equations (\ref{Riccati}) to
exclude the derivatives of $U$. After rather tedious computation
we arrive at \bea \nonumber
\delta\theta_{\a\beta}&=&-\frac{\lambda^2}{\kappa}(\delta_{\a}{}^{\mu}\delta_{\beta}{}^{\nu}-\eps_{\a}{}^{\mu}\eps_{\beta}{}^{\nu})
{\rm Tr} \bigg[
A_{\mu}(A_{\nu}-UA_{\nu}U^{-1})+A_{\mu}(A_{\nu}-UA_{\nu}U^{-1})
\bigg]
\\
 &+&\frac{\lambda(\lambda^2-1)}{\kappa}{\rm
Tr}\big(\eps_{\a}{}^{\mu}[A_{\mu},A_{\beta}]U+\eps_{\beta}{}^{\mu}[A_{\mu},A_{\a}]U\big)\,
. \eea By using this formula we can now find how the stress tensor
varies under the B\"acklund transform \bea \nonumber \delta
T_{\a\beta}&=&\frac{\lambda^2}{\kappa}(\gamma_{\a\beta}\gamma^{\mu\nu}-\delta_{\a}{}^{\mu}\delta_{\beta}{}^{\nu}
+\eps_{\a}{}^{\mu}\eps_{\beta}{}^{\nu}) {\rm Tr} \bigg[
A_{\mu}(A_{\nu}-UA_{\nu}U^{-1})+A_{\mu}(A_{\nu}-UA_{\nu}U^{-1})
\bigg]
\\
 &+&\frac{\lambda(\lambda^2-1)}{\kappa}{\rm
Tr}U\Big(\eps_{\a}{}^{\mu}[A_{\mu},A_{\beta}]+\eps_{\beta}{}^{\mu}[A_{\mu},A_{\a}]
+\gamma_{\a\beta}\eps^{\mu\nu}[A_{\mu},A_{\nu}]\Big)\, . \eea Let
us consider the first term in the expression above. It involves
three tensor structures which are however not independent. Indeed,
there is the \emph{epsilon identity} which reads \bea
\eps^{\mu\nu}\eps_{\a\beta}=\delta_{\a}{}^{\nu}\delta_{\beta}{}^{\mu}-
\delta_{\a}{}^{\mu}\delta_{\beta}{}^{\nu}\, .\eea This identity
implies \bea \eps_{\a}{}^{\mu}\eps_{\beta}{}^{\nu}=
\delta_{\a}{}^{\nu}\delta_{\beta}{}^{\mu}-\gamma_{\a\beta}\gamma^{\mu\nu}\,
. \eea Therefore, \bea
\gamma_{\a\beta}\gamma^{\mu\nu}-\delta_{\a}{}^{\mu}\delta_{\beta}{}^{\nu}
+\eps_{\a}{}^{\mu}\eps_{\beta}{}^{\nu}=\delta_{\a}{}^{\nu}\delta_{\beta}{}^{\mu}-\delta_{\a}{}^{\mu}\delta_{\beta}{}^{\nu}
\, . \eea Since this expression is multiplied by a tensor which is
symmetric under the permutation of the $\mu$ and $\nu$ indices, the
contribution of the term under consideration vanishes. Thus, 
\bea
\nonumber \delta
T_{\a\beta}=\frac{\lambda(\lambda^2-1)}{\kappa}{\rm
Tr}U\Big(\eps_{\a}{}^{\mu}[A_{\mu},A_{\beta}]+\eps_{\beta}{}^{\mu}[A_{\mu},A_{\a}]
+\gamma_{\a\beta}\eps^{\mu\nu}[A_{\mu},A_{\nu}]\Big)\, . \eea
Thus, we see that the compatibility of the  Virasoro constraints  with
the general solution of the B\"acklund transform requires
$\lambda=\pm 1$.
Note that $\delta T_{\a\beta}$ also
vanishes for $\lambda=0$ which must be the case since for this
value of $\lambda$ we trivially have $\tilde{A}=A$.

\section{General Solution of the B\"acklund Equations}

In the previous section, we have found that the new solution,
generated via the B\"acklund transform, satisfies the Virasoro
constraints if and only if the spectral paramter~$\lambda$ is
restricted to be~$\lambda=\pm 1$. Therefore, in the following we only
consider this case. The equation~(\ref{re}) then implies that the
spectral parameter~$x$ has to be in the range~$-1\leq x\leq 1$.

One way of solving the Riccati equation~(\ref{Riccati}) is using the
perturbation method,~i.e. by expanding the variable~$U$ in a power
series around the points~$x=\pm 1$. We present this computation in
appendix~A. Only these perturbative solutions have been so far been
used in the literature, to determine the \emph{local} conservation
laws~\cite{Ogielski:1979hv}.

\medskip

In this section we show how to express the solutions of the Riccati
equation via solutions of the Riemann-Hilbert problem in a
\emph{non-perturbative} manner. The main result is given in
formula~(\ref{U-sol}).  We will furthermore use this result to
establish a simple relation between the B\"acklund charges and the
local integrals of motion generated by the monodromy matrix of the
fundamental linear problem.

\medskip

To obtain solutions of the Riccati equation we employ the
linearization method of~\cite{HSS}. According to this method, a
solution~$U$ of the Riccati equation can be factorized as 
\bea
U(\sigma,\tau)=XY^{-1}\, . 
\eea 
Here the matrices $X$ and $Y$ are obtained by applying to the initial
data, $X_0$ and $Y_0$, an element~$\mathbf{G}$ of the group SU(4,4)
\bea \left(
\begin{array}{c}
X \\ Y
\end{array}
\right)=\mathbf{G}\left(
\begin{array}{c}
X_0 \\ Y_0
\end{array}
\right) \, , ~~~~~~~\Omega=X_0Y_0^{-1} \, ,  \eea so that the initial
value of $U$ is
$U(0,0)=\Omega$. \footnote{Since the Riccati equation is a differential
equation its solutions depend on an integration constant; this
constant is $\Omega$.} of $U$ is
$U(0,0)=\Omega$.

The matrix $\mathbf{G}$ is subject to the following two conditions
\bea \label{cond}
\mathbf{G}^{\dagger}h_1\mathbf{G}=h_1,~~~~~~~~\mathbf{G}^{\dagger}h_2\mathbf{G}=h_2\,
, \eea where $h_1$ and $h_2$ are the following block matrices \bea
\nonumber
 h_1=\left(
\begin{array}{rr}
\mathbb{I} &  0 \\
0 & -\mathbb{I}
\end{array}
\right) \, , ~~~~~~~~~~h_2=\left(
\begin{array}{rr}
0~~ &  \mathbb{I} \\
\mathbb{I}~~ & -2x\mathbb{I}
\end{array}
\right) \, .\eea
The first condition in (\ref{cond}) means that
$\mathbf{G}$ belongs to SU(4,4) and is necessary for $U$
to be unitary. The second condition is equivalent to the requirement
(\ref{char}). The constant matrix $\Omega$ obeys the same
constraints as the matrix $U$, namely,
$$
\Omega^{\dagger}\Omega=\mathbb{I}\, ,
~~~~~~~~~\Omega+\Omega^{\dagger}=2x\mathbb{I}\, .
$$

\medskip

Solving equations (\ref{cond}) we find
\begin{equation}
\label{cs} {\bf G} =\frac{1}{\ell-\bar{\ell}}
\left(\begin{array}{cc}
 \ell \Psi - \bar{\ell} (\Psi^{\dagger})^{-1}  ~~ & (\Psi^{\dagger})^{-1}-\Psi  \\
\Psi -(\Psi^{\dagger})^{-1}~~ &\ell (\Psi^{\dagger})^{-1}
-\bar{\ell} \Psi
\end{array} \right)\, ,
\end{equation}
where $\ell$ is the same complex parameter as in equation (\ref{re}),
{\it i.e.} it is related to the spectral parameter $x$ as  $(\lambda=1)$ \bea
\ell=x-i\sqrt{1-x^2} \, .\eea Note also that $\ell$ is on the unit
circle because $\ell\bar{\ell}=1$. Finally, if the complex matrix
$\Psi$ in equation (\ref{cs}) satisfies the differential equation \bea
\label{standard} \pa_{\a}\Psi=\L_{\a}(\ell)\Psi\, , \eea where by
definition $\L_{\a}(\ell)$ is the $\L$-operator (\ref{Lc}) with
the following coefficients $a$ and $b$ \bea \label{standardc}
a=\frac{1}{1-\ell^2}\, , ~~~~~~~b=\frac{\ell}{1-\ell^2} \, ,\eea
then the Riccati equation for $U$ is satisfied. We will refer to
equation (\ref{standard}) as \emph{the fundamental linear problem}.

\medskip

To verify the last statement we first find the
matrices\footnote{Note the conjugation rule:
$\Psi^{\dagger}(\ell)\Psi(\bar{\ell})=\mathbb{I}$.} $X$ and $Y$
\bea \nonumber X&=&\big[\ell \Psi(\ell)-\bar{\ell}
\Psi(\bar{\ell}) \big]\Omega +\Psi(\bar{\ell})-\Psi(\ell) \, ,\\
\label{solXY} Y&=&\big[\Psi(\ell) -
\Psi(\bar{\ell})\big]\Omega+\ell \Psi(\bar{\ell})- \bar{\ell}
\Psi(\ell) \, , \eea where the matrix $\Psi$ is normalized as \bea
\Psi(0,0)=\mathbb{I}\, . \eea Next we note the following two
identities valid for the spectral parameter $\ell$ on a circle
\bea
\L_{\a}(\ell)&=&\frac{\eps_{\alpha}{}^{\beta}-\bar{\ell}\delta_{\alpha}{}^{\beta}}{\ell-\bar{\ell}}\Big(\L_{\beta}(\ell)
-\L_{\beta}(\bar{\ell})\Big)\, , \nonumber
\\
\nonumber
\L_{\a}(\bar{\ell})&=&\frac{\eps_{\alpha}{}^{\beta}-\ell\delta_{\alpha}{}^{\beta}}{\ell-\bar{\ell}}\Big(\L_{\beta}(\ell)
-\L_{\beta}(\bar{\ell})\Big)\, . \eea
These identities together with equation (\ref{standard}) for
$\Psi$ are used to obtain the system of evolution equations for
$X$ and $Y$: \bea
\pa_{\a}X&=&\frac{\eps_{\alpha}{}^{\beta}
}{\ell-\bar{\ell}}
\Big(\L_{\beta}(\ell)
-\L_{\beta}(\bar{\ell})\Big)X-\frac{1}{\ell-\bar{\ell}}
\Big(\L_{\alpha}(\ell) -\L_{\alpha}(\bar{\ell})\Big)Y \, ,
\nonumber
\\[1ex]
\label{XY}
\pa_{\a}Y&=&\frac{\eps_{\alpha}{}^{\beta}-2x\delta_{\alpha}{}^{\beta}}{\ell-\bar{\ell}}
\Big(\L_{\beta}(\ell)
-\L_{\beta}(\bar{\ell})\Big)Y+\frac{1}{\ell-\bar{\ell}}
\Big(\L_{\alpha}(\ell) -\L_{\alpha}(\bar{\ell})\Big)X\, .
 \eea
In writing these formulae the following relation has been used
\bea
(\bar{\ell}\Psi(\ell)-\ell\Psi(\bar{\ell}))\Omega+\ell^2\Psi(\bar{\ell})-\bar{\ell}^2\Psi(\ell)
=-X+2x Y\, . \eea Now the Riccati equation for $U$ easily follows
from the system (\ref{XY}).

\medskip

\noindent Thus, the solution for $U$ reads as
\bea
\label{U-sol}
U=\Big[\ell \Psi(\ell)\big(\Omega- \bar{\ell}\big) +
\bar{\ell}\Psi(\bar{\ell})\big(\Omega-\ell \big) \Big]
\Big[\Psi(\ell)\big(\Omega- \bar{\ell}\big) -
\Psi(\bar{\ell})\big(\Omega-\ell \big) \Big]^{-1}\, .\eea Note
that the matrices $\Omega-\ell$ and $\Omega-\bar{\ell}$ are not
invertible. As was already mentioned, since
$$
\ell\bar{\ell}=1, ~~~~~~~~\ell+\bar{\ell}=2x\, ,
$$
the variables $\ell$ and $\bar{\ell}$ are the eigenvalues of
$\Omega$ and
$$
(\Omega-\bar{\ell})(\Omega^{\dagger}-\bar{\ell}) =0, ~~~~~~~~~
\Omega-\ell=-(\Omega^{\dagger}-\bar{\ell}),~~~~~(\Omega-\bar{\ell})(\Omega-\ell)=0\,
.
$$
These properties allow us to define two hermitian (and orthogonal)
projectors \bea {\sl
\Omega}^+=\frac{\Omega-\bar{\ell}}{\ell-\bar{\ell}}\, , ~~~~~~
{\sl \Omega}^-=-\frac{\Omega-\ell}{\ell-\bar{\ell}} \eea which
provide an orthogonal decomposition of the identity: $\Omega^+
 +\Omega^-={\mathbb
I}$,~ $\Omega^{\pm}\Omega^{\mp}=0$.

\section{Matching the B\"acklund and Monodromy Charges}

We would now like to establish a connection between the
conservation laws generated by the B\"acklund transform and the
conservation laws arising in the standard monodromy approach.
Our starting point is the nonperturbative solution (\ref{Riccati}) of the
B\"acklund equation, and the expression for the B\"acklund current~(\ref{simple}).
Let us start by noting the following important relation \bea
\label{wichtig}
\gamma^{\a\beta}\L_{\beta}\left(\frac{1}{1-x^2},\frac{x}{1-x^2}\right)
=\frac{2}{\ell-\bar{\ell}}\eps^{\a\beta}\Big(\L_{\beta}(\ell)-\L_{\beta}(\bar{\ell})\Big)
. \eea
Here on the left hand side, the Lax operator $\L$ is the same as in
equation (\ref{Lc}), with the coefficients $a$ and $b$ parametrized by
$x$. This is this Lax operator which determines the conserved
B\"acklund current (\ref{simple}).~The Lax operator which appears on
the right hand side in (\ref{wichtig}) is the same as $\L(\ell)$ which
defines the fundamental linear problem (\ref{standard}). The
coefficients $a$ and $b$ of the operator $\L(\bar{\ell})$ are given by
equations (\ref{standardc}) with the obvious substitution $\ell\to
\bar{\ell}$.
\medskip

Using the equation~(\ref{XY}) for~$Y$, we thus obtain \bea
\pa_{\a}YY^{-1}=
\frac{\eps_{\alpha}{}^{\beta}-2x\delta_{\alpha}{}^{\beta}}{\ell-\bar{\ell}}
\Big(\L_{\beta}(\ell)
-\L_{\beta}(\bar{\ell})\Big)+\frac{1}{\ell-\bar{\ell}}
\Big(\L_{\alpha}(\ell) -\L_{\alpha}(\bar{\ell})\Big)U \, .
\eea
Taking the trace of this equation, we arrive at \bea {\rm
Tr}\big(\pa_{\a}YY^{-1}\big)= {\rm Tr}\Big(\frac{\L_{\alpha}(\ell)
-\L_{\alpha}(\bar{\ell})}{\ell-\bar{\ell}}\Big)U\, .\eea
Therefore, by using equation (\ref{wichtig}) for the current
(\ref{simple}), we find the simple expression
\bea J^{\alpha}=\gamma^{\alpha\beta}{\rm
Tr}\L_{\beta}(x)U=2\eps^{\alpha\beta}{\rm
Tr}\Big(\frac{\L_{\beta}(\ell)
-\L_{\beta}(\bar{\ell})}{\ell-\bar{\ell}}\Big)U=2\eps^{\alpha\beta}{\rm
Tr}\big(\pa_{\beta}YY^{-1}\big)\, . \eea
In this form, the  conservation
of the current is obvious. The current is topological and thus is
conserved without using the equations of motion for the fundamental
fields. However, the corresponding charge is conserved if and only
if the current is a periodic function of~$\sigma$. Periodicity of
$J^{\alpha}$ then imposes certain restrictions on the initial
value $\Omega$. To better understand this issue we compute the
charge (\ref{QJ}), \bea \pi\mathbf{Q}(x)={\rm Tr}\log\big(
Y(2\pi,\tau)Y(0,\tau)^{-1}\big)\, .\eea Differentiating this
expression with respect to $\tau$ and using equations (\ref{XY}) we find that it
is indeed time-independent provided
$$
X(2\pi,\tau)Y(2\pi,\tau)^{-1}=X(0,\tau)Y(0,\tau)^{-1}
$$
or, in other words, that the B\"acklund solution is periodic:
$U(2\pi)=U(0)$.
\medskip

Let us now study the periodicity property of the B\"acklund
solution in more detail. Clearly, periodicity of $U$ is
equivalent to the following requirement
\bea X(2\pi,\tau)&=&X(0,\tau){\rm M} \nonumber \\
Y(2\pi,\tau)&=&Y(0,\tau){\rm M} \, , \eea {\it i.e.}~the matrices $X$ and
$Y$ have to have the same monodromy ${\rm M}$. By using
equations~(\ref{solXY}),~it is easy to see that this requirement is
equivalent to  \bea
\Psi(0,\tau)^{-1}\Psi(2\pi,\tau)\Omega^{+}=\Omega^+{\rm
M} \, ,\nonumber \\
\label{RM} \bP(0,\tau)^{-1}\bP(2\pi,\tau)\Omega^{-}=\Omega^-{\rm
M}\, , \eea
where we have introduced the concise notation
$\Psi\equiv\Psi(\ell)$ and $\bar{\Psi}\equiv \Psi(\bar{\ell})$.
There are several important facts following from these equations.

First, let us note that these equations make sense because
both $\Psi(0,\tau)^{-1}\Psi(2\pi,\tau)$ and
$\bP(0,\tau)^{-1}\bP(2\pi,\tau)$ are time-independent as
straightforwardly follows from the evolution equation for $\Psi$.
Thus, $\rm M$ is also time-independent. Introducing the monodromy
${\rm T}(\tau)$ for the solution $\Psi$ \bea \label{CM}
\Psi(2\pi,\tau)={\rm T }(\tau)\Psi(0,\tau)\,  \eea we can express
our basic $\tau$-independent quantities via the value of the
corresponding monodromy matrix\footnote{Note that the monodromy
${\rm T}$ do depend on $\tau$, only its spectral invariants are
conserved.} at $\tau=0$ as $\Psi(0,\tau)^{-1}\Psi(2\pi,\tau)={\rm
T}(0)$ and $\bP(0,\tau)^{-1}\bP(2\pi,\tau)=\bar{{\rm T}}(0)$.
Second, by adding the equations (\ref{RM}) we obtain \bea {\rm M}={\rm
T}(0)\Omega^+ +\bar{\rm T}(0)\Omega^- \, .\eea Furthermore, we derive from
equations (\ref{RM})  the equations which  determine $\Omega^\pm$: \bea
\label{cOmega} \Omega^-{\rm T }(0)\Omega^{+}=0\, , ~~~~~~~
\Omega^+\bar{\rm T}(0)\Omega^{-}=0\, . \eea As a side remark,  note
that the determinants of both $\Psi$ and $\bar{\Psi}$ are $\tau$-
and $\sigma$-independent and therefore that
$\det\Psi=\det\bar{\Psi}=1$ (since this is the case at the initial
point $\tau=\sigma=0$). As a consequence, $\det {\rm
T}=\det\bar{\rm T}=1$.

\medskip

Let us now diagonalize $\Omega$ with some unitary matrix $h$ so
that $\Omega^{\pm}$ take the block-form
\bea h\Omega^{+}h^{-1}=\left(\begin{array}{cc} \mathbb{I} ~&~ 0 \\
0 ~&~ 0
\end{array}\right)\, , ~~~~~~~~
h\Omega^{-}h^{-1}=\left(\begin{array}{cc} 0 ~&~ 0 \\ 0 ~&~
\mathbb{I}
\end{array}\right)\, .
\eea
Then according to equations (\ref{cOmega}) we see that the matrices
$h^{-1}{\rm T}(0)h$ and $h^{-1}\bar{\rm T}(0)h$ must have the
(block) lower and upper triangular structure respectively \bea
\label{2mon} {\mathbf T}=h^{-1}{\rm T}(0)h =
\left(\begin{array}{cc} {\mathbf T}_1 ~&~ {\mathbf T}_2 \\
0 ~&~ {\mathbf T}_4
\end{array}\right)\, , ~~~~~~
\bar{\mathbf T}=h^{-1}\bar{\rm T}(0)h =
\left(\begin{array}{cc} \bar{\mathbf T}_1 ~&~ 0 \\
\bar{\mathbf T}_3 ~&~ \bar{\mathbf T}_4
\end{array}\right)\, \, .
\eea This allows one to write the conserved charge in the
following factorized form \bea \label{factor}
\pi\mathbf{Q}(x)={\rm Tr}\log {\rm M}={\rm Tr}\log
\mathbf{T}_1+{\rm Tr}\log\bar{\mathbf{T}}_4\, . \eea
The reality property of the B\"acklund charge  implies the
conjugation rule according to which $\mathbf{T}_1^{\dagger}$ is
related to $ \bar{\mathbf{T}}_4$ by a similarity transformation.

\medskip

Let us denote by $\exp(ip_k(\ell))$, where  $k=1\ldots 4$, the
eigenvalues of the monodromy matrix ${\rm T}$. The function
$p_k(\ell)$ is known as the quasi-momentum or the Floquet
function. An important property of the quasi-momentum is that it
generates \emph{local} integrals of motion upon expansion around
the poles of the Lax connections \cite{FT}, which in our case are
at~\mbox{$\ell=\pm 1$}.

\begin{figure}[t]
\psfrag{l}{$\ell$} \psfrag{lb}{$\bar\ell$} \psfrag{exp}{\vbox{\hbox{\small
expansion point}\hbox{\small for $\lambda =1$}}}
\psfrag{exp2}{\vbox{\hbox{\small expansion point}\hbox{\small for $\lambda= -1$}}}
\psfrag{Im(l)}{\!\!\!${\rm Im}\,\ell$} \psfrag{Re(l) = X}{${\rm
Re}\,\ell=x$} \psfrag{1}{1} \psfrag{-1}{-1}
\begin{center}
\includegraphics*[width=.6\textwidth]{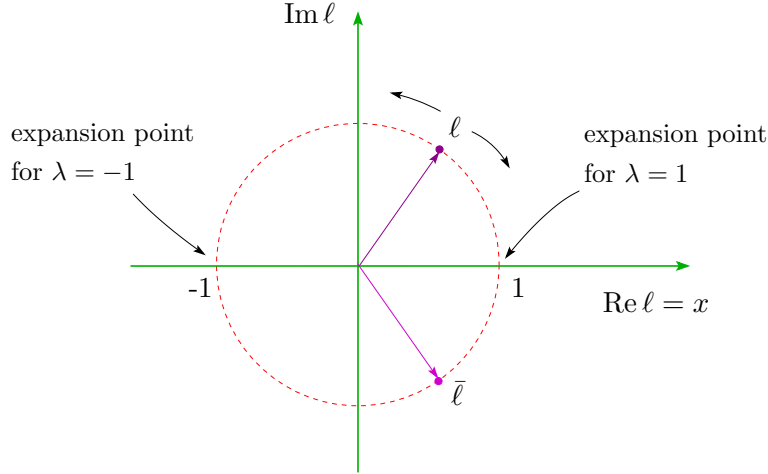}
\end{center}
\caption{The B\"acklund generating function $\mathbf{Q}(x)$ for
$x\in [-1, 1]$ is given by the sum of the quasi-momenta in the
upper and lower half-planes (semi-circles).}
\end{figure}

\medskip

Finally, it remains to note that the triangle monodromies
$\mathbf{T}$ and $\mathbf{\bar{T}}$ in equations (\ref{2mon}) can
be brought to diagonal form by corresponding similarity
transformations and their spectra coincide with that of ${\rm T}$
and $\bar{\rm T}$ respectively. In this way we have established
the following remarkably simple relation between the local charges
generated by the B\"acklund transform \cite{AS} and their cousins
arising in the conventional monodromy approach \cite{KMMZ}, \bea
\label{finBasic}
\mathbf{Q}(x)=\frac{i}{\pi}\big(p_1(\ell)+p_2(\ell)+p_3(\bar{\ell})+p_4(\bar{\ell})
\big) \, .\eea Here $\sum_{k=1}^4 p_k(\ell)=0$ and the spectral
parameter $\ell$ of the linear problem (\ref{standard}) is related
to the spectral parameter $x$ of the B\"acklund transform as \bea
\ell=x-i\sqrt{1-x^2} \, .\eea We stress that our derivation does
not require any gauge fixing and the result is valid for an
arbitrary world-sheet metric $\gamma$. Also taking the $\log$'s in
equation (\ref{factor}) we assumed that all $p_i$'s are on the principle
branch of the $\log$.

\medskip

Equation (\ref{finBasic}) should be understood in a perturbative
sense when the left and the right hand side admit a well-defined
asymptotic expansion around~$x=\ell=\bar{\ell}=1$. Of course, the
same relation is true for the second series of the conservation
laws upon expanding around $x=\ell=\bar{\ell}=-1$.

\medskip
So far our discussion was quite general and applied to the
principal sigma-model. To carry over this construction to the coset
model describing strings on \adss one has to find an embedding of
the coset into the group SU(2,2)$\times$SU(4) that is compatible
with additional constraints like equation (3.4). This is also needed to
guarantee that the form of the coset element is preserved under
the B\"acklund transformations. In appendix B we show that the
embedding of \adss into SU(2,2)$\times$SU(4) described in section
2 obeys these compatibility requirements. This allows us to
conclude that the same formula (\ref{finBasic}) remains valid for
the sphere part of the coset model. For the AdS sector the
matching formula~(\ref{finBasic}) looks the same provided $p_k(x)$
are quasi-momenta related to the AdS monodromy.

\medskip Finally we note that the quasi-momenta are defined up to
permutations. On the other hand the formula (\ref{finBasic}) does
not seem to be permutation invariant. For the coset model in
question it is known \cite{AF} that \emph{at leading order} in the
perturbative expansion around a pole the quasi-momenta exhibit a
degenerate behavior: they all coincide up to a sign, two of them
are positive and the other two are negative. Fixing up $p_1(\ell)$
we then define $p_2(\ell)$ to be such a quasi-momentum for which
$-i\log\det {\rm T}_1=p_1(\ell)+p_2(\ell)$ is non-zero at leading
order in $1/(1-x)$ expansion. In the next section we will check
equation (\ref{finBasic}) for the first two orders in the
perturbative expansion.

\section{Monodromy vs. B\"acklund for rigid strings}

Here we would like to check the basic formula (\ref{finBasic}) by
explicitly comparing the few leading charges arising in the expansion
of the generating function for B\"acklund charges~$\mathbf{Q}(x)$,
with that of the quasi-momentum~$p(\ell)$. In general, finding the
higher charges from the mondoromy is rather involved; however we make
progress by computing their values on certain string
configurations. In particular, the rigid string
solutions~\cite{AFRT,ART} provide an excellent tool for probing the
higher hidden charges~\cite{AS}.

\medskip
We choose to work with a solution which describes a  rigid string with a
circular profile, and  carrying two non-vanishing spins in the
five-sphere. This solution can be conveniently written in terms of
the standard Jacobi elliptic functions as follows \cite{AFRT} \bea
\nonumber
&&q^1+iq^2=\mbox{sn}(a\sigma, \st )\exp(iw_1\tau)\, , \\
\label{sol}
&&q^3+iq^4=\mbox{cn}(a\sigma, \st)\exp(iw_2\tau)\, ,\\
\nonumber &&q^5=q^6=0. \eea Here $w_{12}^2= w^2_1 - w^2_2 $ is
related to the elliptic modulus $\st$ through the closed string
periodicity condition:
 \bea a\equiv \sqrt{\frac{w_{12}^2}{\st}}=
\frac{2}{\pi}\ellK(\st)\, , \eea where $\ellK$ is the complete
elliptic integral of the first kind. The modulus $\st$ can be
further expressed via the~${\rm S}^5$ spins but we do not need
this here.

\medskip
The generating function for the B\"acklund charges ${\cal
E}(\gamma)$ on rigid string solutions was obtained in \cite{AS}.
In this work another spectral parameter denoted by $\gamma$ has
been used (not to be confused with our definition of the
world-sheet metric). It is related to $x$ we use here by \bea
\label{rel} \gamma^2= \frac{1-x}{1+x} \, .\eea The generating
function $\mathbf{Q}(x)$ is obtained from the generating function
${\cal E} (\gamma)$ of \cite{AS} by multiplying it  with a certain
factor, namely, \bea \mathbf{Q}(x)=-4
(1-x)^{-\frac{3}{2}}\sqrt{1+x} ~{\cal E}(\gamma)\, , \eea where
the spectral parameter $\gamma$ is related to $x$ through
(\ref{rel}). Using these relations we then extract from the
results of \cite{AS} the following asymptotics
\bea \label{bcl}
\mathbf{Q}(x)\stackrel{x\to
1}{=}\frac{2\sqrt{2}}{\sqrt{1-x}}Q_{-1}+\frac{1}{\sqrt{2}}Q_1\sqrt{1-x}+\ldots\,
, \eea where all the coefficients $Q_k$ are functions of $\ellK$
and $\ellE$ -- the elliptic integrals of the first and second kind
respectively. In particular, \bea Q_{-1}&=&{\cal
E }\ , \\
Q_{1}&=&{\cal E} -\frac{32}{\pi^2\cE}\ellK(t)\ellE(t)
-\frac{64(t-1)}{\pi^4\cE^3}\ellK(t)^4 \, .\eea Here ${\cal
E}=\sqrt{\frac{1}{t}(w_{1}^2+(t-1)w_2^2)}$ is the space-time
energy of the string.

\medskip
Now let us consider  the monodromy (\ref{CM}) of the fundamental
linear problem related to the solution (\ref{sol}). Due to
$q_5=q_6=0$ the $\su(4)$ Lax connection $\L$ can be split by an
appropriate (constant) similarity transformation into two
independent $\su(2)$ connections~$\L^{\pm}$. Moreover, the time
dependence of the latter is trivially factored out as
\bea \L_{\a}
\to
\left(\begin{array}{cc} \mathscr{R}_{+}\L^{+}_{\a}\mathscr{R}_{+}^{\dagger} & 0 \\
0 & \mathscr{R}_{-}\L^{-}_{\a}\mathscr{R}_{-}^{\dagger}
\end{array}\right)\, .
 \eea
Here
\bea \mathscr{R}_{\pm}= \left(\begin{array}{cc} e^{i\frac{w_1\pm w_2}{2}\tau} & 0 \\
0 & e^{-i\frac{w_1\pm w_2}{2}\tau}
\end{array}\right)\, \eea
and the $\su(2)$ matrices $\L^{\pm}_{\a}$ are time-independent. In
particular, the $\sigma$-components of $\L^{\pm}$ read \bea
\nonumber {\footnotesize
\hspace{2mm}~\L^+_{\sigma}(\ell)=\frac{1}{1-\lle^2}
\left(\begin{array}{cc} i\ell(w_1\sn^2 a\sigma-w_2\cn^2 a\sigma) &
-a\dn a\sigma-i\ell (w_1+w_2)\sn a\sigma \cn a\sigma             \\
a\dn a\sigma-i\ell(w_1+w_2)\sn a\sigma \cn a\sigma
     &   -i\ell(w_1\sn^2 a\sigma-w_2\cn^2 a\sigma)
\end{array}\right)   } \, ,
\\
\nonumber {\footnotesize
\hspace{2mm}~\L^-_{\sigma}(\ell)=\frac{1}{1-\lle^2}
\left(\begin{array}{cc} i\ell(w_1\sn^2 a\sigma+w_2\cn^2 a\sigma) &
-a\dn a\sigma-i\ell (w_1-w_2)\sn a\sigma \cn a\sigma             \\
a\dn a\sigma-i\ell(w_1-w_2)\sn a\sigma \cn a\sigma
     &   -i\ell(w_1\sn^2 a\sigma+w_2\cn^2 a\sigma)
\end{array}\right)   } \, .
\eea Clearly they just differ by the substitution $w_2\to -w_2$.
These connections are used to construct the corresponding
monodromies \bea {\rm
T}_{\pm}(\lle)=\;\stackrel{\longleftarrow}{\exp}\int_0^{2\pi} {\rm
d}\s \L^{\pm}_{\s}(\lle) \, .\eea Furthermore we define the following
$\sigma$-dependent matrices \bea {\rm
T}_{\pm}(\sigma)=\;\stackrel{\longleftarrow}{\exp}\int_0^{\sigma}
{\rm d}\s \L^{\pm}_{\s}(\lle) \,  \eea which are solutions to the
differential equations \bea \pa_{\sigma}{\rm
T}_{\pm}=\L^{\pm}_{\sigma}{\rm T }_{\pm}\, . \eea In what follows
we will discuss ${\rm T}\equiv {\rm T}_{-}$, the results for ${\rm
T}_+$ are obtained by the substitution $w_2\to -w_2$.
\medskip

Let us represent ${\rm T}(\sigma)$ as \bea {\rm
T}(\s)=g(\s)\mathscr{D}(\s)g^{-1}(0),
~~~~~~~~~~~\mathscr{D}(\s)=\exp(id(\s)\s_3) \, ,\eea where $g(\s)$
is a \emph{periodic unitary} gauge transformation.\footnote{The
procedure we use here is equivalent to diagonalizing the Lax
connection around one of its poles by an appropriate regular
unitary gauge transformation. } Thus, the trace of the monodromy
satisfies ${\rm Tr}\,{\rm T}(2\pi)=2\cos d(2\pi)$ which implies that
the quasi-momentum $p(\ell)=d(2\pi)$ is \bea
p(\ell)=\frac{1}{2}\arccos {\rm Tr} {\rm
T}(2\pi)=\int_0^{2\pi}{\rm d}\s \pa_{\s}d(\s)\, . \eea Here the
last formula is a consequence of $d(0)=0$.
Introducing the parametrization \bea
g=\frac{1}{\sqrt{1+\rho\bar{\rho}}} \left(\begin{array}{rr}
1 ~&~ \rho \\
-\bar{\rho} ~&~ 1
\end{array}
\right)\, , ~~~~~~~~~~~~~~ \L_{\sigma}=\left(\begin{array}{cc}
iu & v \\
-\bar{v} & -iu
\end{array}
\right) \eea one finds that the differential equation for ${\rm
T}(\sigma)$ boils down to the following system \bea \label{1}
\pa_{\s}\rho &=& v+2iu \rho +\bar{v} \rho^2\,, \\
\pa_{\s}d&=&u+\frac{1}{2i}(\rho\bar{v}-v\bar{\rho})\,. \eea In
particular, the first equation is of the (scalar!) Riccati type.
Then the generating function for the string integrals of motion is
given by \bea p(\lle)=\int_0^{2\pi}d\s\Big[
u+\frac{1}{2i}(\rho\bar{v}-v\bar{\rho}) \Big]\,. \eea To solve equation
(\ref{1}) we assume the expansion \bea
\rho=\rho_0+(1-\lle)\rho_1+\ldots \eea around the pole  of $\L$ at
$\lle=1$. In particular, the solution for $\rho_0$  is \bea
\rho_0=-i\frac{w_2+\cE+(w_1-w_2)\sn^2(a\s,t)}
{a\dn^2(a\s,t)-i(w_1-w_2)\sn(a\s,t)\cn(a\s,t)} \, .\eea

\medskip
\noindent Around $\ell=1$ the quasi-momentum is expanded as \bea
\frac{p(\ell)}{\pi}=\frac{1}{\ell-1}p_{-1}+p_0+(\ell-1)p_1+\ldots
\eea Performing rather involved integrations we have found a few
leading charges. They are
\bea \nonumber
p_{-1}&=&\cE \, ,\\
\nonumber p_0&=&\frac{1}{2}\cE +\frac{1}{\cE}(\cE-w_1)(\cE-w_2)
-\frac{2(\cE-w_1)}{\ellK(t)}\Pi\left(\frac{w_2-w_1}{w_2-\cE},t\right)\, , \\
\label{pc}
p_1&=&-\frac{1}{4}\cE+\frac{4}{\pi^2\cE}\ellK(t)\ellE(t)
+\frac{8(t-1)}{\pi^4\cE^3}\ellK(t)^4 \, ,\\
\nonumber p_2&=&\frac{1}{8}\cE-\frac{2}{\pi^2\cE}\ellK(t)\ellE(t)
-\frac{4(t-1)}{\pi^4\cE^3}\ellK(t)^4 \\
\nonumber
&-&\frac{w_1w_2}{\cE^2}\Big[\frac{2}{\pi^2\cE}\ellK(t)\ellE(t)
+\frac{8(t-1)}{\pi^4\cE^3}\ellK(t)^4\Big] \, .\eea Here the
frequencies $w_{1,2}$ are expressed via the space-time energy of
the string \bea
w_1=\frac{1}{\pi}\sqrt{\pi^2\cE^2+4(t-1)\ellK(t)^2}\, , ~~~~~~~~~
w_2=\frac{1}{\pi}\sqrt{\pi^2\cE^2-4\ellK(t)^2}\,  \eea and $\Pi$
stands for the standard elliptic integral of the third kind.

\medskip
Some comments are in order. The charge $p_{-1}$ is independent of
$w_2$ and therefore it is the same for both monodromies ${\rm
T}_{\pm}$. This means that the quasi-momenta exhibit a degenerate
behavior at leading order in the $1/(\ell-1)$ expansion \cite{AF}.
Furthermore we note that the charge $p_0$ is rather distinguished from
the rest as it is the only one which contains the elliptic integral of
the third kind. Most importantly, this charge is not invariant under
$w_2\to -w_2$. Therefore, the degeneracy of the quasi-momenta observed
at leading order gets removed. This is an important fact because it
allows one to treat the quasi-momenta~$p_k(\ell)$ with the
corresponding pole part subtracted as analytic functions associated to
different sheets of a unique Riemann surface \cite{KMMZ}. To observe
the splitting of the eigenvalues of the monodromy, unitarity of the
gauge transformation diagonalizing the $\L$-operator around its pole
is essential.

\medskip

Finally, we compute \bea \nonumber
i\left(\frac{p(\ell)}{\pi}-\frac{p(\bar{\ell})}{\pi}\right)
=\frac{\sqrt{2}}{\sqrt{1-x}}p_{-1}+\frac{1}{2\sqrt{2}}(-p_{-1}-8p_1)\sqrt{1-x}+\ldots
\eea Note that the charge $p_0$ does not appear in this expansion.
Since $p_{-1}$ and $p_{1}$ are independent of $w_2$ the
contribution of the quasi-momenta associated to ${\rm T}_+$ will
be the same at this order. Therefore, \bea \label{check}
\frac{i}{\pi}\left(p_1(\ell)+p_2(\ell)-p_1(\bar{\ell})-p_2(\bar{\ell})\right)
=\frac{2\sqrt{2}}{\sqrt{1-x}}p_{-1}+\frac{1}{\sqrt{2}}(-p_{-1}-8p_1)\sqrt{1-x}+\ldots
\eea Now substituting here the expressions (\ref{pc}) for the $p's$ we
observe that equation (\ref{check}) perfectly reproduces the first two
terms in the expansion (\ref{bcl})! This provides another
non-trivial check of our basic formula (\ref{finBasic}).

\section*{Acknowledgments}
We would like to thank Fernando Alday,  Sergey Frolov, Andrei
Mikhailov, Matthias Staudacher and Arkady Tseytlin for interesting
discussions. The work of G.~A. was supported in part by the
European Commission RTN programme HPRN-CT-2000-00131 and by RFBI
grant N02-01-00695.

\appendix

\section{Perturbative Solution of the B\"acklund Equations}

Here we discuss the perturbative solutions of the B\"acklund equations
which allow one to determine the \emph{local} conservation laws of the
model. Our treatment can be viewed as the matrix generalization of the
vector approach of~\cite{Ogielski:1979hv}. We also assume $\lambda=\pm
1$ so that the B\"acklund solutions satisfy the Virasoro constraints.

\medskip
To start, we write the matrix $U$ in the form \bea \label{B1}
U=\frac{x}{\lambda}\mathbb{I}+\bf{P} \, ,\eea where ${\bf P}$ is
anti-hermitian, ${\bf P}^{\dagger}+{\bf P}=0$, and obeys the
condition \bea \label{p} {\bf P}^{\dagger}{\bf
P}=\mathbb{I}-\Big(\frac{x}{\lambda}\Big)^2=-{\bf P}^2\, . \eea
Thus, the matrix ${\bf P}$ has degenerate eigenvalues which are
$\pm i\sqrt{1-\frac{x^2}{\l^2}}$. In terms of ${\bf P}$ the
Riccati equation acquires the form \bea \nonumber \kappa {\bf P
}_{\a}&=& \label{pert} \frac{2x}{\lambda}(x^2-\l^2)
A_{\a}+(1+\lambda^2-2x^2)[A_{\a},{\bf P }] -2x\lambda {\bf P
}A_{\a}{\bf
P } +\\
\nonumber
 &+& \epsilon_\alpha{}^\beta
\Big(\frac{1}{\lambda}(x^2-\l^2)(1+\lambda^2)A_{\beta}+x(1-\l^2)
[A_{\beta},{\bf P }] -\lambda(1+\lambda^2){\bf P }A_{\beta}{\bf P
} \Big)\, . \eea Equivalently this can be cast into the form
\begin{equation}
{\bf P}_\alpha = {1 \over  1 + \l^2} [{\mathscr L}_
\alpha-x\epsilon_\alpha{}^\beta {\mathscr{L}}_\beta -\half \lambda
\epsilon_\alpha{}^\beta [{\bf P},{\mathscr{L}}_\beta ] , {\bf P }]
 \, ,
\end{equation}
where we made use of equation (\ref{p}). One can easily check that this
equation is compatible with the symmetry properties of ${\bf P}$.
In fact, it can be viewed as the matrix differential equation for
an element ${\bf P}=\tilde{q}^iq^j\Gamma_{ij}$ of $\su(4)$.

\medskip
From equation (\ref{B1}) we see that for $\lambda=1$ the matrix $U\to
\mathbb{I}$ when $x\to 1$
 while for $\lambda=-1$ we will have the
same asymptotic behavior provided $x\to -1$. This shows that for
$\lambda=1$ the Riccati equation will have a well-defined
perturbative expansion around $x=1$ and for $\lambda=-1$ the
expansion must be concentrated around $x=-1$.
\medskip

Let us define $\zeta=\pm\sqrt{1-x^2}$ where $``+"$ is for
$\lambda=1$ and $``-"$ is for $\lambda=-1$. It is convenient to
introduce a rescaled matrix $\mathscr{P}$:
 \bea \label{exp}
\mathscr{P}=\frac{{\bf P}}{\zeta}
=\sum_{n=0}^{\infty}\zeta^n\mathscr{P}_n \, \eea obeying the
condition $\mathscr{P}^{\dagger}\mathscr{P}=\mathbb{I}$. For
$\lambda=\pm 1$ the Riccati equation boils down to \bea
\zeta(-2\mathscr{P}_{\a}+[A_{\a},\mathscr{P}])=\pm
\sqrt{1-\zeta^2}A_{\a}+\eps_{\a}{}^{\beta}A_{\beta}+
\mathscr{P}(\pm
\sqrt{1-\zeta^2}A_{\a}+\eps_{\a}{}^{\beta}A_{\beta})\mathscr{P} \,
. \eea Now one can realize that upon substituting here the
expansion (\ref{exp}) we will get recurrent relations which would
allow us to solve for $\mathscr{P}_n$ in terms of lower
coefficients \mbox{$\mathscr{P}_k$, $k<n$}. Most importantly, we see that
in the perturbative treatment the original differential problem
was replaced by an algebraic one. Thus, finding solutions does
not involve integration and, as a consequence, the solution appears
to be a local function of the fields and their derivatives.

\medskip

As an example, let us find explicitly the leading term
$\mathscr{P}_0$. We get \bea \label{lead1} {}[A_{\a}\pm
\eps_{\a}{}^{\beta}A_{\beta}, \mathscr{P}_0 ] = 0\, ,
\eea together with the condition
\begin{equation}
\label{lead2} {\mathscr P}_0^2 = -\mathbb{I} \,
\end{equation}
that follows from the linearized expression (\ref{p}). Recall that
the reference gauge connection~$A_{\alpha}$ equals
\begin{equation}
A_{\alpha} = g_{\alpha} g^{-1} = q_{\alpha}^i q^j \Gamma_{ij} \, .
\end{equation}
One can show that equations (\ref{lead1}), (\ref{lead2}) have the
following solutions
\begin{equation}
\label{Pcal0} {\mathscr P}_0  = {2\over || q_{\alpha}^{\pm}|| }
(q_\alpha^i \pm \epsilon_\alpha{}^\beta q_\beta^i) q^j
\Gamma_{ij}\, ~~~~~{\mbox or}~~~~~{\mathscr P}_0  = -{2\over ||
q_{\alpha}^{\pm}|| } (q_\alpha^i \pm \epsilon_\alpha{}^\beta
q_\beta^i) q^j \Gamma_{ij}\, ,
\end{equation}
where there is no summation over the index $\alpha$. Here we use
the notation \bea q_{\alpha}^\pm =\big(\delta_{
\alpha}{}^{\beta}\pm\eps_{ \alpha}{}^{\beta}\big) q_{\beta} \,
,~~~~~~~|| q_{\alpha}^\pm ||^2 \equiv q_{\alpha}^\pm  \cdot
q_{\alpha}^\pm .\eea Since \bea q_{\pm \sigma}=\pm \frac{1\pm
\gamma_{\tau\sigma}}{\gamma_{\tau\tau}}q_{\pm \tau} \eea one finds
that \bea \label{propr} {q^\pm_{\tau} \over ||q^{\pm}_\tau||}
={q^\pm_{\sigma} \over ||q^{\pm}_\sigma||} ={q^{\pm \tau} \over
||q^{\pm \tau}||} = { q^{\pm \sigma} \over ||q^{\pm \sigma}||} \,
. \eea Now using the properties (\ref{propr}) we can see that
(\ref{Pcal0}) manifestly satisfies the relations (\ref{lead1}) and
(\ref{lead2}).

\section{Coset model}
Introduce the following six unitary $4 \times 4$ matrices
\begin{eqnarray}
\nonumber \Gamma_1&=&{\footnotesize\left(
\begin{array}{cccc}
  0 & 1 & 0 & 0 \\
  -1 & 0 & 0 & 0 \\
   0 & 0 & 0 & 1 \\
   0 & 0 & -1 & 0
\end{array} \right)},\hspace{0.3in}\Gamma_2={\footnotesize\left(
\begin{array}{cccc}
  0 & i & 0 & 0 \\
  -i & 0 & 0 & 0 \\
   0 & 0 & 0 & -i \\
   0 & 0 & i & 0
\end{array} \right)},\hspace{0.3in}\Gamma_3={\footnotesize\left(
\begin{array}{cccc}
  0 & 0 & -1 & 0 \\
  0 & 0 & 0 & 1 \\
   1 & 0 & 0 & 0 \\
   0 & -1 & 0 & 0
\end{array} \right)}, \\
\nonumber \Gamma_4&=&{\footnotesize \left(
\begin{array}{cccc}
  0 & 0 & -i & 0 \\
  0 & 0 & 0 & -i \\
   i & 0 & 0 & 0 \\
   0 & i & 0 & 0
\end{array} \right)},\hspace{0.3in}~\Gamma_5={\footnotesize \left(
\begin{array}{cccc}
  0 & 0 & 0 & 1 \\
  0 & 0 & 1 & 0 \\
   0 & -1 & 0 & 0 \\
   -1 & 0 & 0 & 0
\end{array} \right)},\hspace{0.3in}\Gamma_6={\footnotesize \left(
\begin{array}{cccc}
  0 & 0 & 0 & -i \\
  0 & 0 & i & 0 \\
   0 & -i & 0 & 0 \\
   i & 0 & 0 & 0
\end{array} \right)}.
\end{eqnarray}
These matrices satisfy the algebra
$$
\Gamma_i\Gamma_j^{\dagger}+\Gamma_j\Gamma_i^{\dagger}=2\delta_{ij}\,
.
$$
An embedding $g$ of a coset element describing the five-sphere
into SU(4) is conveniently described as \bea g=q_i\Gamma_i\, ,\eea
where the coordinates $q_i$ satisfy the constraint $q_iq_i=1$.
Note that the six $\Gamma$-matrices above are antisymmetric.
Therefore, the coset element is obtained by intersecting the
unitarity condition $g^{\dagger}g=1$ with the requirement
$g^{t}=-g$.

Let us define
$\Gamma_{ij}=\half(\Gamma_i\Gamma_j^{\dagger}-\Gamma_j\Gamma_i^{\dagger})$.
The matrices $\Gamma_{ij}$ obey the following algebra \bea \{
\Gamma_{ij},\Gamma_{kl} \}=
2(\delta_{il}\delta_{jk}-\delta_{ik}\delta_{jl}) \, ,\eea \bea
 [ \Gamma_{ij},\Gamma_{kl} ]=2(
\delta_{jk}\Gamma_{il}+\delta_{il}\Gamma_{jk}-\delta_{ik}\Gamma_{jl}-\delta_{jl}\Gamma_{ik})
\, , \eea {\it i.e.} $\Gamma_{ij}$ generate the $\su(4)$ algebra.
Another way to represent the $\su(4)$ generators is to use
$\bar{\Gamma}_{ij}=\half(\Gamma_i^{\dagger}\Gamma_j-\Gamma_j^{\dagger}\Gamma_i)$.
One can easily see that $\bar{\Gamma}_{ij}=(\Gamma_{ij})^*$
corresponds to the anti-chiral representation of $\su(4)$.

The element $U$ is then \bea
U=\half\tilde{q}_iq_j(\Gamma_i\Gamma_j^{\dagger}+\Gamma_j\Gamma_i^{\dagger})
+\half\tilde{q}_iq_j(\Gamma_i\Gamma_j^{\dagger}-\Gamma_j\Gamma_i^{\dagger})
=(\tilde{q}q)\mathbb{I}+\tilde{q}_iq_j \Gamma_{ij} \, ,\eea where
we have introduced $\Gamma_{ij}=\half(
\Gamma_i\Gamma_j^{\dagger}-\Gamma_j\Gamma_i^{\dagger})$. If we
define $\chi=\lambda U$, then equation (\ref{pchi}) acquires the form
\bea \label{spb} \lambda U+\bar{\lambda}U^{\dagger}=2x\mathbb{I}\,
. \eea Upon substitution of the coset element $U$ equation (\ref{spb})
reduces to \bea \nonumber
&& (\l+\bl)(\tilde{q}q)=2x \\
\nonumber && \l \Gamma_{ij}+\bl \Gamma_{ij}^{\dagger}=0\, ,
~~~i\neq j\, .
 \eea
Since the matrices $\Gamma_{ij}$ are anti-hermitian we have to
require that $\l=\bl$ and, therefore, for the scalar product
$(\tilde{q}q)$ we obtain $(\tilde{q}q)=\frac{x}{\l}$.
\medskip

For completeness we also provide a similar representation for
$\su(2,2)$ and describe the AdS sector of the model. Consider the
following metric
$$\eta_{ij}={\rm diag}(1,1,1,1,-1,-1)\, $$
and the matrix $E$: \bea {\rm E}={\rm diag}(1,1,-1,-1)\, . \eea
Once again we introduce the six $4\times 4$ matrices
\begin{eqnarray}
\nonumber \Gamma_1&=&{\footnotesize\left(
\begin{array}{cccc}
  0 & 0 & -i & 0 \\
  0 & 0 & 0 & i \\
   i & 0 & 0 & 0 \\
   0 & -i & 0 & 0
\end{array} \right)},\hspace{0.3in}\Gamma_2={\footnotesize\left(
\begin{array}{cccc}
  0 & 0 & 1 & 0 \\
  0 & 0 & 0 & 1 \\
   -1 & 0 & 0 & 0 \\
   0 & -1 & 0 & 0
\end{array} \right)},\hspace{0.3in}\Gamma_3={\footnotesize\left(
\begin{array}{cccc}
  0 & 0 & 0 & i \\
  0 & 0 & i & 0 \\
   0 & -i & 0 & 0 \\
   -i & 0 & 0 & 0
\end{array} \right)}, \\
\nonumber \Gamma_4&=&{\footnotesize \left(
\begin{array}{cccc}
  0 & 0 & 0 & 1 \\
  0 & 0 & -1 & 0 \\
   0 & 1  & 0 & 0\\
   -1 & 0 & 0 & 0
\end{array} \right)},\hspace{0.25in}~\Gamma_5={\footnotesize \left(
\begin{array}{cccc}
  0 & 1 & 0 & 0 \\
  -1 & 0 & 0 & 0 \\
   0 & 0 & 0 & 1 \\
   0 & 0 & -1 & 0
\end{array} \right)},\hspace{0.3in}\Gamma_6={\footnotesize \left(
\begin{array}{cccc}
  0 & i & 0 & 0 \\
  -i & 0 & 0 & 0 \\
   0 & 0 & 0 & i \\
   0 & 0 & -i & 0
\end{array} \right)}.
\end{eqnarray}
These matrices obey the following algebra \bea \Gamma^i{\rm
E}\Gamma^{j\dagger}+ \Gamma^j{\rm
E}\Gamma^{i\dagger}=-2\eta^{ij}{\rm E} \, .\eea If we introduce
$g=p^i\Gamma^i$ we then see that this element satisfies \bea
\label{group} g^{\dagger}{\rm E}g={\rm E} \, ,\eea provided $p^i$
obey $\eta_{ij}p^ip^j=-1$. Equation  (\ref{group}) defines the group
SU(2,2). We further introduce \bea \Gamma_{ij}=\half(\Gamma_i{\rm
E }\Gamma_j^{\dagger}{\rm E}-\Gamma_j{\rm E}\Gamma_i^{\dagger}{\rm
E})\, ,~~~~~~\Gamma_{ij}^{\dagger}=-{\rm E}\Gamma_{ij}{\rm E}.
\eea These matrices obey the commutation relations of the
$\su(2,2)$ algebra \bea [ \Gamma_{ij},\Gamma_{kl} ]=
2(\eta_{ik}\Gamma_{jl}-\eta_{jk}\Gamma_{jl}+\eta_{jl}\Gamma_{ik}-\eta_{il}\Gamma_{jk})
\, \eea and also \bea \{ \Gamma_{ij},\Gamma_{kl} \}=
2(\eta_{il}\eta_{jk}-\eta_{ik}\eta_{jl}) \, .\eea The AdS element
$U=\tilde{g}g^{-1}$ is then represented as \bea \label{cosetads}
U=-(\tilde{p}p) +\tilde{p}^ip^j\Gamma_{ij}\, , \eea where
$(\tilde{p}p)=\eta_{ij}p^ip^j$. It is easy to see that the AdS
analogue of equation (\ref{spb}) is \bea \label{lat} \lambda{\rm E}
U+\bar{\lambda}U^{\dagger}{\rm E}=2x{\rm E}\, . \eea This equation
is compatible with the coset element (\ref{cosetads}) provided
$\lambda$ is real and $(\tilde{p}p)=-\frac{x}{\lambda}$. Further
we note that equation (\ref{lat}) can be rewritten as \bea \label{st}
U+U^{-1}=2\frac{x}{\lambda}\mathbb{I}\, . \eea Thus, being written
in terms of $U$ and $U^{-1}$ equations (\ref{char}) and (\ref{st}) look
the same and, therefore, lead to the same form of the Riccati
equation (\ref{Riccati}).

\end{document}